\documentclass[article]{elsarticle}

\usepackage{lineno}
\journal{Journal of \LaTeX\ Templates}

\begin{document}

\begin{frontmatter}

\title{New opacity measurement principle for LMJ-PETAL laser facility}

\author{M. Le Pennec \fnref{mymainaddress}}
\author{X. Ribeyre \fnref{mysecondaryaddress}}
\author{ J.-E. Ducret \fnref{mymainaddress,mysecondaryaddress}}
\author{ S. Turck-Chi\`{e}ze\fnref{mymainaddress}}


\address[mymainaddress]{Service d'Astrophysique, CEA/DSM/IRFU, CE Saclay, 91190 Gif-sur-Yvette, FRANCE}
\address[mysecondaryaddress]{Centre Lasers Intenses et Applications CELIA, UMR 5107 CNRS-CEA-Universit\'{e} de Bordeaux, UMR 5107, 351, cours de la Lib\'{e}ration, F-33405 Talence, FRANCE}

\begin{abstract}
Stellar seismology reveals some interior properties of thousands of solar-type stars but the solar seismic sound speed stays puzzling since a decade as it disagrees with the Standard Solar Model (SSM) prediction. One of the explanations of this disagreement may be found in the treatment of the transport of radiation from the solar core to the surface.
As the same framework is used for other stars, it is important to check precisely the reliability of the interacting cross sections of photons with each species in order to ensure the energy transport for temperature T $>$ 2 - 10$^{6}${~}K and density  $\rho$ $>$ 0.2  g/cm$^{3}$. In this paper, we propose a new technique to reach the domain of temperature and density found in the solar radiative interior. This technique called the Double Ablation Front (DAF) is based on a high conversion of the laser energy into X-rays thanks to moderated Z material irradiated by laser intensity between $\rm 1.5{~}\times{~}10^{15}{~}W/cm^{2}$ and $\rm 4{~}\times{~}10^{15}{~}W/cm^{2}$. This high conversion creates, in addition to the electronic front a second ablation front in the moderated Z material. Between the two fronts there is a plateau of density and temperature that we exploit to heat a sample of iron or of oxide. The first simulations realized with the hydrodynamic code CHIC show that this technique allows to reach conditions equivalent to half the radiative zone of the Sun with high stability both in time and space. We examine the possibility to measure both iron and oxygen absorption spectra.
\end{abstract}


\end{frontmatter}

%\linenumbers

\section{Introduction}

\begin{figure}
\begin{center}
 \includegraphics[scale=0.46]{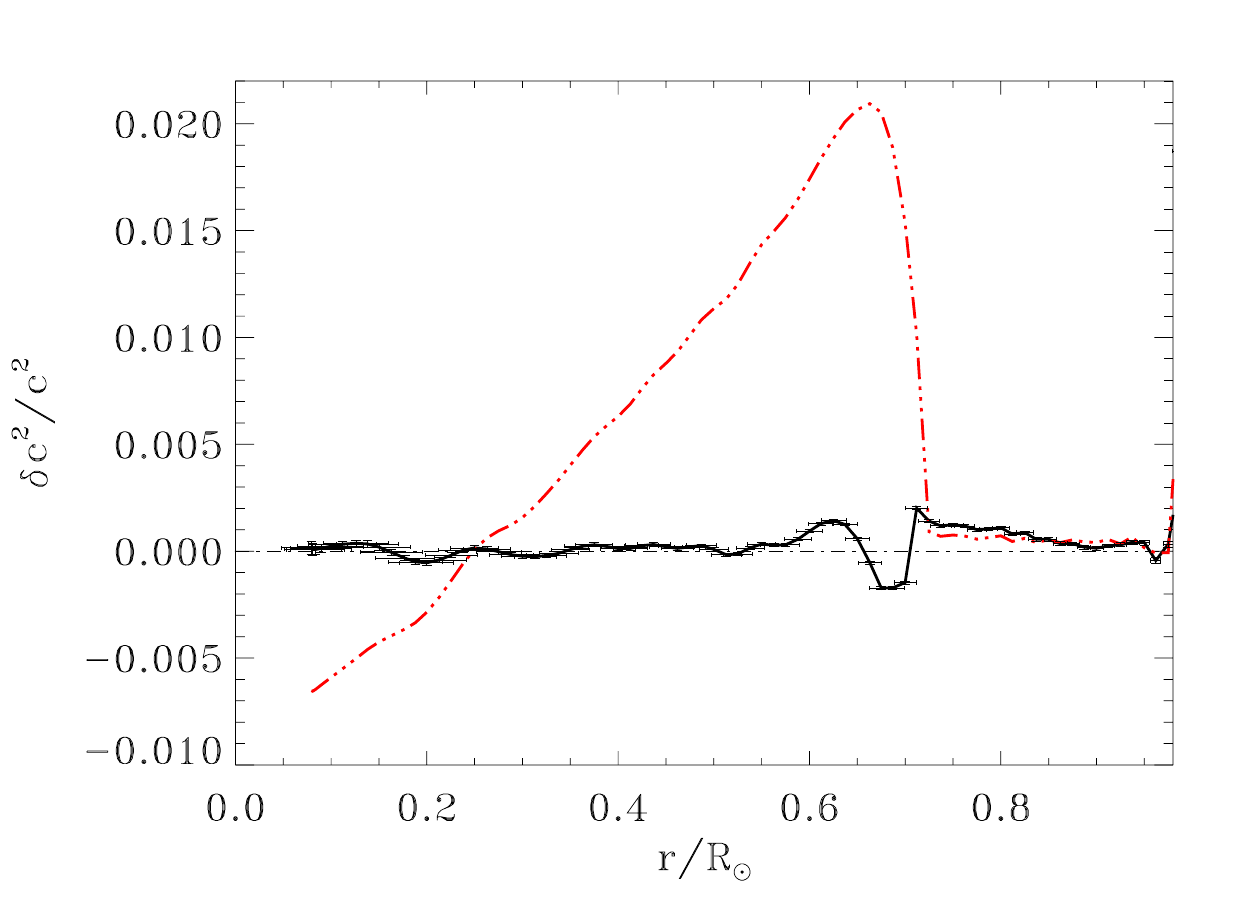} 
 \caption{Relative differences between the squared sound speed coming from seismology (SoHO) and from the SSM (red). A seismic model has been calculated with the same equations than SSM but it has been adjusted to respect as properly as possible the seismic results. Its relative differences with observations are drawn with a black continuous line, the seismic error bars are shown on that model. Adapted from Ref. \cite{Turck2011} (Color online).}
   \label{fig1}
   \end{center}
\end{figure}

The SSM including the updated photospheric composition \cite{Asplund2009} in carbon, oxygen and nitrogen (CNO) disagrees with helioseismic radial profiles and neutrino detections \cite{Turck2004,Turck2011,TurckCouvidat, Basu2014}. This is particularly visible on the sound speed profile that is sensitive to the detailed internal solar composition through the opacity coefficients \cite[and references therein]{Turck1997}: Figure \ref{fig1} recalls the relative difference between the squared sound speed coming from seismology and from the SSM. 

This discrepancy between the observed sound speed and the sound speed predicted by the SSM varies largely greater than the vertical error bar \cite{Turck2011, Basu2014}. To explain this difference, three hypotheses, which could all exist simultaneously, have been advanced \cite{Turck2011,Turck2009}. One can first question the transfer of energy from the center of the Sun to the surface through the opacity coefficients: the atomic calculations in plasma conditions could underestimate the Rosseland mean opacities which directly drive the radiative transfer. The second idea is to put some doubt on the gravitational settling which could be underestimated for CNO and heavy elements due to an incorrect treatment of the radiative acceleration of elements towards the surface with, as a consequence, incorrect central abundances. The last hypothesis concerns the energetic balance equation: the Sun could produce slightly more energy  ($< $ 5{~}\%) than it liberates at the surface and this additional energy would be transformed into macroscopic motions existing in the radiative zone \cite{Turck2010}, so the energetic balance of the SSM is incomplete as it does not contain any internal dynamical effects. 

Determining the origin of this discrepancy (in the opacity ingredients or in some limitation of the solar model) would be an important step toward a better understanding of the solar interior. As the two first hypotheses put in question the way the radiative transfer is used and as the third hypothesis could be difficult to verify except by a very precise neutrino detection, it is important to test the opacity calculations in details. So, this situation requires an experimental validation of the radiative transfer calculations at the conditions of the solar radiative zone (see Table \ref{tab1} for the solar internal conditions to be explored). 

\section{An experimental challenge}
\begin{figure}
 \begin{center}
 \includegraphics[scale=0.6]{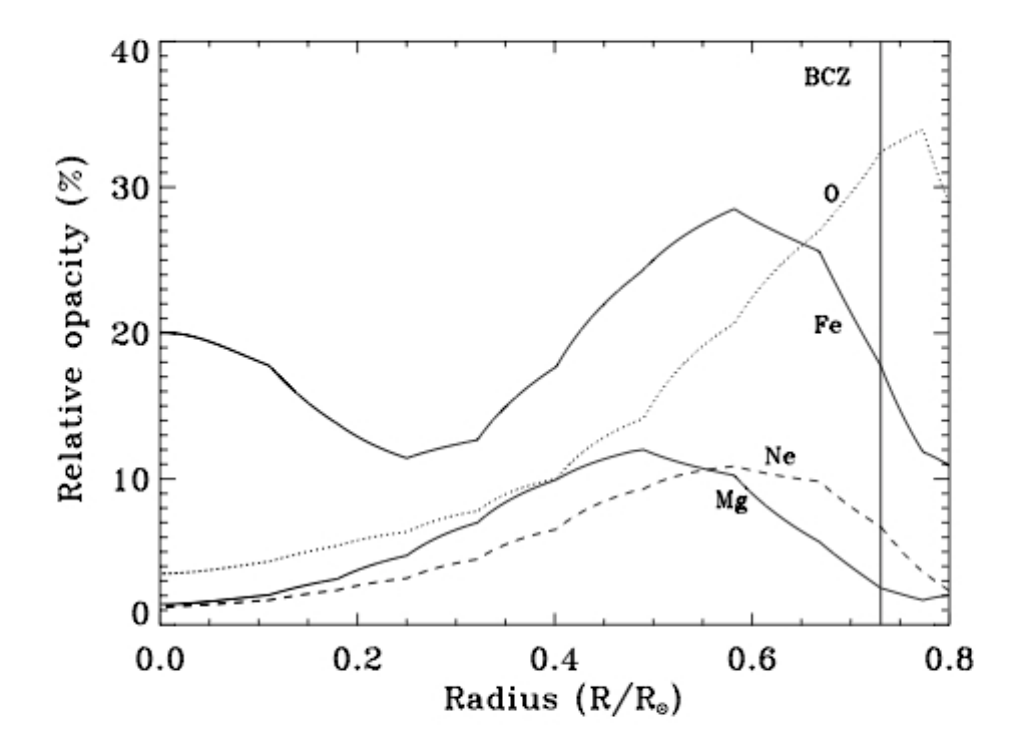} 
 \caption{Relative contribution of the most important heavy element to the total Rosseland opacity (including H and He) for the internal conditions of the Sun, the composition of Ref. \cite{Asplund2009}, using OPAL opacities \cite{Iglesias1996}. From Ref. \cite{Turck2010}.}
   \label{fig2}
   \end{center}
\end{figure}

\begin{figure}[h!]
%\begin{center}
 \includegraphics[scale=0.16]{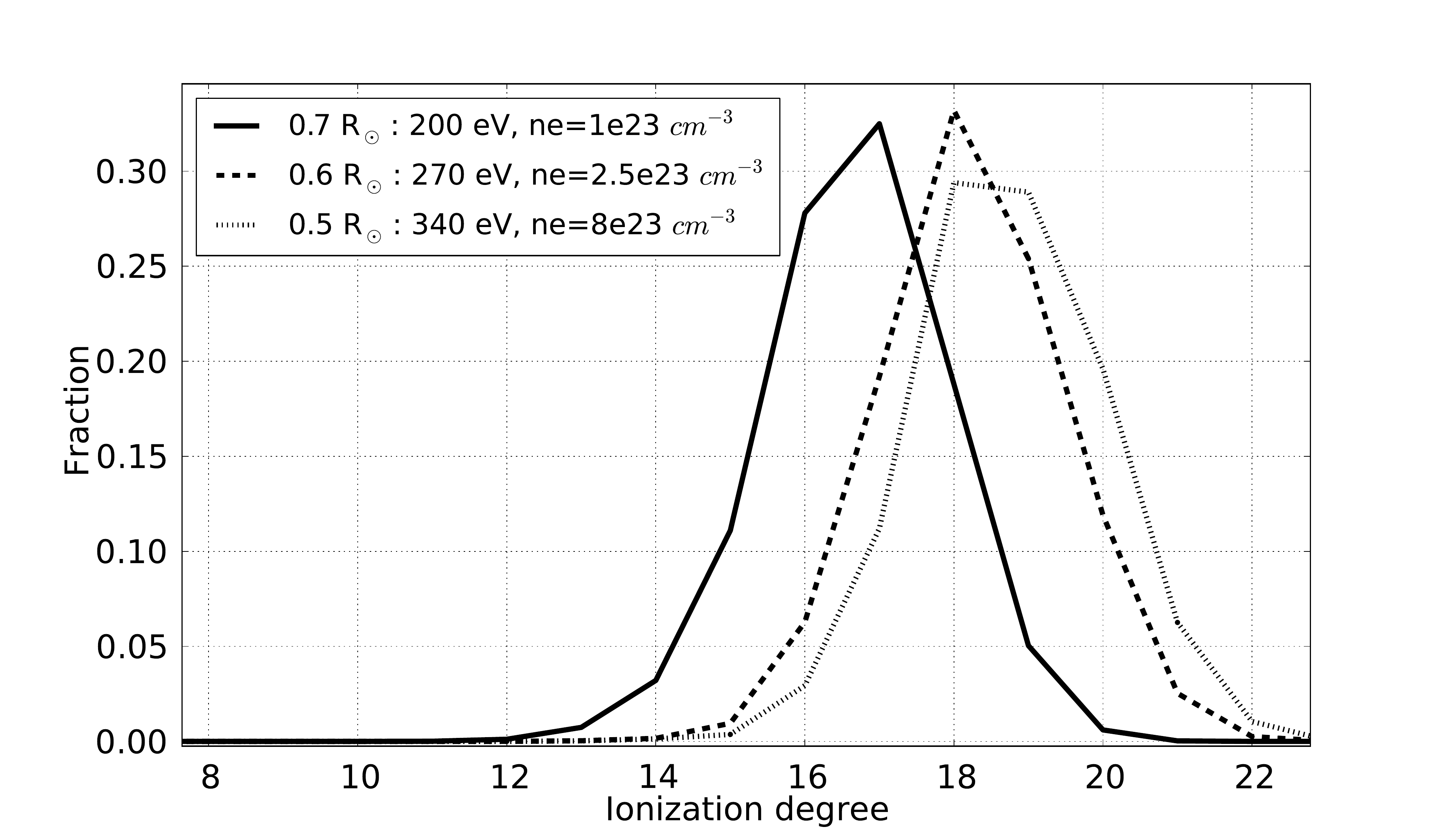} 
 \includegraphics[scale=0.16]{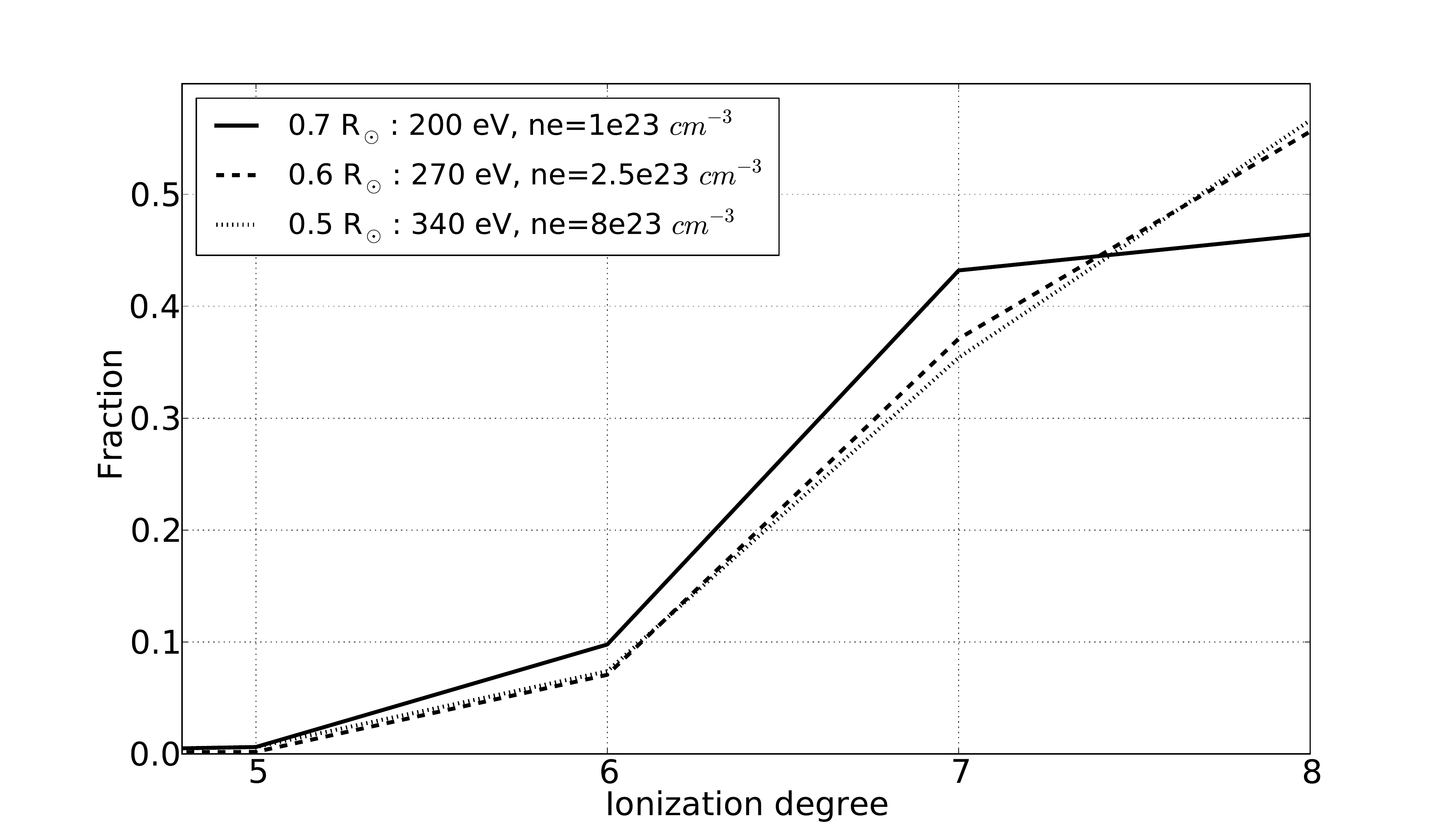} 
 \caption{Charge state distributions of iron (left) and oxygen (right) at 0.5 (...), 0.6 (-{}-{}-) and 0.7 (full line) R$_\odot$ obtained with FLYCHK \cite{Chung2005}.}
   \label{fig3}
%   \end{center}
\end{figure}

\begin{figure}
\begin{center}
 \includegraphics[scale=0.20]{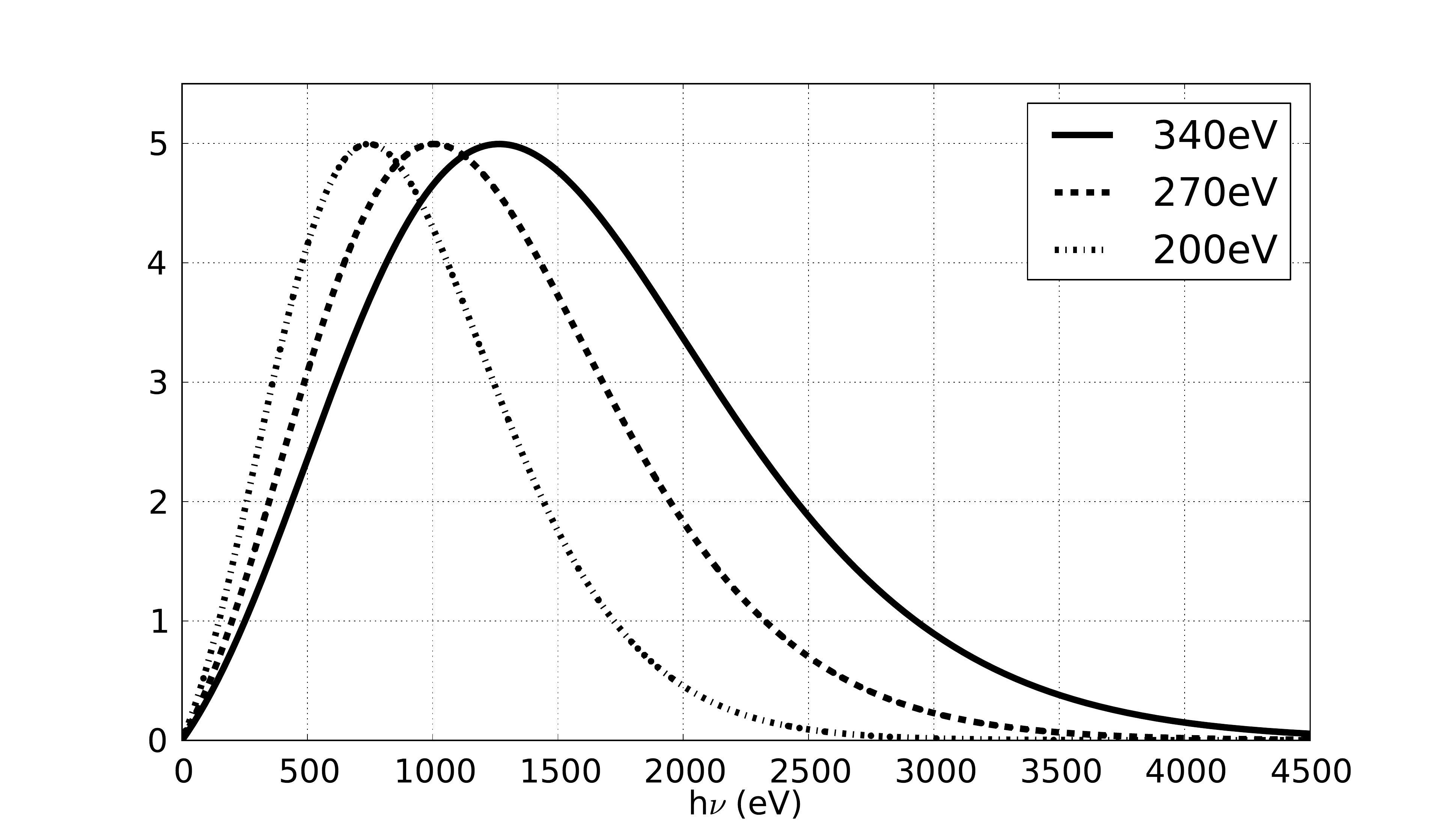} 
 \caption{Rosseland mean ponderation at the three different temperatures.}
   \label{fig4}
   \end{center}
\end{figure}

The Sun is principally constituted of hydrogen and helium which are, in almost the whole Sun completely ionized. However, even though heavy elements are present only at a few percents in mass (iron represents only about 10$^{-3}$ of the hydrogen contribution in mass fraction), they contribute significantly to the global opacity \cite{Turck2010}. \\ 
Figure \ref{fig2} represents the respective role of the main elements contributing to the global opacity along the solar radial profile. The elements heavier than $\mathrm{^{4}He}$ represent only 1.4{~}\% in mass fraction but their contribution to the opacity coefficient is up to 70{~}\% at the basis of the convective zone. The largest contribution of iron and oxygen in the radiative zone is due to their bound-bound contributions. Iron is never completely ionized in the Sun: even in the center, it still contributes to opacity with bound-bound and bound-free transitions. Its contribution to the total opacity is around 20 - 30{~}\% in most of the radiative zone. Oxygen, the third element in abundance, changes from fully to partly ionized above 0.4 R$_{\odot}$ and plays a major role to trigger the convection instability around 0.7 R$_{\odot}$. Unfortunately, these plasma properties have never been verified in laboratory nor the absorption energy spectra.
Several problems have made such measurements difficult. To properly evaluate the opacity of the considered elements, one has to reproduce their charge state distribution and the free-electron density N$_{e}$ at the targeted conditions. As shown in Table \ref{tab1}, the free-electron densities in the radiative zone are $10^{23}-10^{24}{~}\mathrm{cm^{-3}}$.  Then, one needs to determine monochromatic single-element opacity in well diagnosed plasmas with good uniformity of temperature and density. 
\begin{table}[h]
\caption{\label{tab:table1} Summary of solar conditions  found in Ref. \cite{Couvidat2003} and in MESA solar model \cite{Paxton2011}}
\begin{center}
\begin{tabular}{cccc}
Solar radius (r/R$_\odot$)&T (eV)& $\rho$ (g.cm$^{-3}$) & N$_e$ (cm$^{-3}$)\\
\hline
\hline
0.5 & 340 & 1.36 & 8 $\times$ 10$^{23}$\\
0.6 & 270 & 0.50 & 2.5 $\times$ 10$^{23}$\\
0.7 & 200 & 0.21 & 1 $\times$ 10$^{23}$
\end{tabular}
\label{tab1}
\end{center}
\vspace{-0.5cm}
\end{table}
Figure \ref{fig3} shows the charge state distributions of iron and oxygen at the conditions of Table 1. As iron is partially ionized, with a great number of bound electrons, the calculation of its opacity is challenging. \\
\linebreak In the solar radiative zone, the radiation transport is described by a diffusion approximation using the Rosseland mean opacity $\kappa_R$: 
\begin{center}
$$
\frac{1}{\kappa_R} = \frac{\int{d \nu \frac{1}{\kappa(\nu)} \frac{dB}{dT}}}{\int{d\nu \frac{dB}{dT}}} 
$$
\end{center}
where B is the Planck function, T is the temperature and $\kappa(\nu)$, the spectral opacity. The shape of the weighting function {dB}/{dT} is represented on Figure \ref{fig4} for the same three conditions. This Rosseland mean ponderation directly determines the spectral range of interest: the maximum is around to {h$\nu$}/{kT}~$\simeq$4, it determines the part of the spectrum which contributes the most to opacity. So, for the solar conditions of Table \ref{tab1}, the range of interest is principally between 500 and 2500 eV as previously shown by \cite{Bailey2009}. This gives precious information for the qualification of diagnostics.

\section{State of the art}
A first experiment was performed at the Z-pinch facility of Sandia National Laboratory at T$_{e}$ = 156 $\pm$ 6 eV and N$_{e}$ = 6.9 $\pm$ 1.7 $\times$ 10$^{21}$ cm$^{-3}$ where the transmission of a mixed Mg and Fe plasma was measured  \cite{Bailey2007}. Recently, the same group performed another experiment with the same principle also on a mixed Mg and Fe sample and they reached T$_{e}$ = 196 $\pm$ 6 eV and N$_{e}$ = 3.8 $\pm$ 0.8 $\times$ 10$^{22}$ cm$^{-3}$ \cite{Nagayama2014}. These conditions are not so far from the solar ones but there is a clear discrepancy between experiment and all the opacity calculations up to now. 

Other experiments involving lasers and compared to theoretical opacities \cite{Winhart1996, Popovics2000, Turck2013} were limited to T$_{e}$ $<$ 100 eV and density $\rho$  $<$ 10$^{-2}${~}g/cm$^{-3}$, which is far from solar conditions. To heat material at T$_e$$>$200 eV at relatively high density, there are different methods with laser beams:  with a shock (giving T$_e$ of few tens eV and $\rho$  $\sim$ 3{~}-{~}4 times the solid density, with small gradients) \cite{Sawada2009}, with a thermal wave (T$_e$ $\sim$ few tens eV and density around the solid density, with high gradients) \cite{Sawada2009} and finally with short pulse lasers (T$_e$ $\sim$ 600 - 800 eV and $\rho{~}>$ solid density) \cite{Hoarty2010,Hoarty2013}. There is also another technique, proposed in Ref. \cite{Keiter2013}, using X-rays conversion in holhraum, which is presently in development. They hope to obtain T$_e$ of some hundreds eV and $\rho$ around 0.1 the solid density.

We propose in this paper another approach, called the Double Ablation Front (DAF) to limit the gradients in the foil. This method is promising for measuring iron absorption spectrum in solar conditions. We explain in the next sections the interest of this technique and show some characteristics of the design that we apply here to an oxide measurement. This DAF has been experimentally put in evidence by \cite{fujioka2004} and extensively studied for fusion by \cite{Sanz2009} and \cite{Drean2010}.

\section{CHIC code}
We have used the CHIC code  (Code d'Hydrodynamique et d'Implosion du CELIA) \cite{Breil2007} for our simulations.
This hydrodynamic code is Lagrangian and dedicated to ICF calculations. It operates in 1D and 2D axially symmetric geometries and includes three-dimensional ray tracing for laser beam propagation. The laser energy release is modeled by inverse bremsstrahlung absorption. The electronic thermal transport is described in the classical Spitzer-H\"{a}rm approximation, with a flux limitation of 0.06 \cite{Malone1975,Delettrez1986}. In our calculations, the wavelength of the laser beams is set to 0.351 $\mu$m corresponding to both LMJ and NIF beams, the diameter of the spot being 400 $\mu$m. The pulse duration is 2.2 ns with a linear ramp of 0.2 ns at the beginning and at the end of the pulse. The SESAME table \cite{SESAME} is used for the equation of state and we converge in number of groups for the treatment of the radiative transport through the diffusion approximation. In the following simulations, we did not take into account any magnetic field generation nor a non local conduction.

\section{Double Ablation Front}
The DAF approach is based on an enhancement of the radiative effects in the target, due to a high conversion between the laser energy and X-rays in a moderated Z material.    
\begin{figure*}[h!]
\begin{center}
 \includegraphics[scale=0.50]{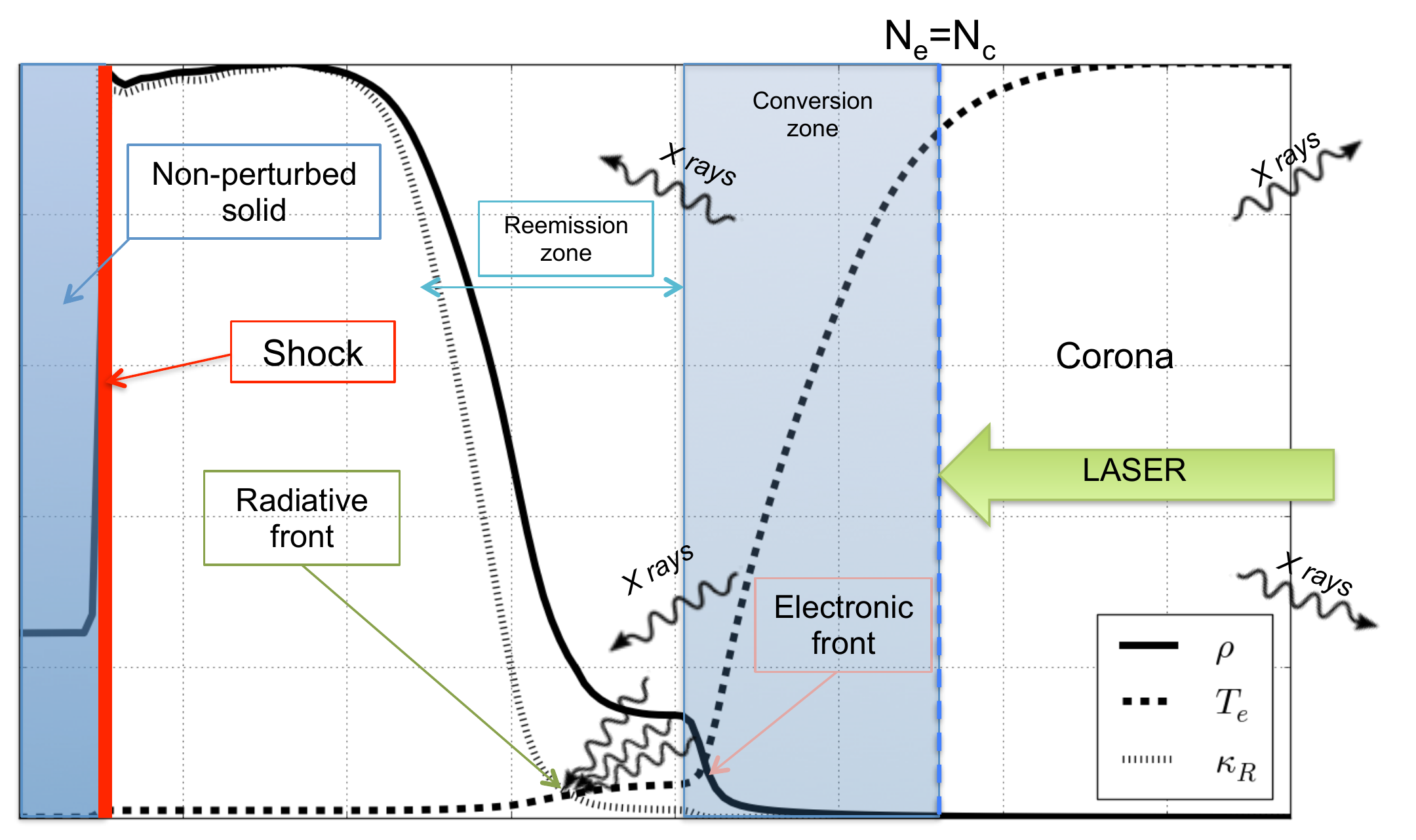} 
\caption{Schematic profiles of electronic temperature $T_e$, density $\rho$ and opacity $\kappa_R$ at a given time in a layer of moderated Z ablator. One sees different regions: the corona (N$_{e}$$<$ $N_{c}$; $T_{e}$ around 3 keV and $\rho$ $\sim$ 0.02 g/cc); the conversion zone (T$_e$ around 1 keV and low density); the reemission zone (T$_e$ of several hundreds eV and $\rho$ $\sim$ 1 g/cc), heated by the X-rays \cite{Marshak1958}; the shock wave and the non-perturbed solid. Two ablation fronts appear, one due to electrons (electronic front), the other due to photons (radiative front) (Color online).}
   \label{fig5}
   \end{center}
\end{figure*}
The laser when deposing its energy near the critical density $N_c$ strongly heats this region. X-rays are emitted, through several processes (principally Bremsstrahlung emission (free-free transitions), emission by electron-ion recombination (bound-free transitions) and line emission (bound-bound transitions), dominant in the case of high Z materials). 
These X-rays are isotropically emitted:  the radiation propagating toward the target in high-density region ($\sim$ solid density) and relatively low temperature ($\sim$ hundreds eV) will be highly absorbed because of the high opacity of this region, creating a radiative ablation front \cite{Sanz2009, Drean2010}. The higher the atomic number Z is, the more important will be the X-rays emission (Bremsstrahlung emission proportional to Z$^{3}$). 
At the same time, the electrons also transport the energy through the target, beyond the critical density. The electronic temperature decreasing in the denser region, they deposit their energy, creating a thermal ablation front (see Figure \ref{fig5}). 
Between the two fronts, there is a "plateau" of density which can be exploited as presented in the following section. Because of the difference of velocities of the two fronts, this plateau region extends with time, ensuring therefore its stability for diagnosis during an experiment. The diffusion approximation is satisfying in this ablation region because of its high opacity which reduces drastically the photon mean free path. This hypothesis is no longer valid in the low optical depth region of the corona but it has already been shown that the diffusion approximation yields to quantitatively good results \cite{Yu2007}. \\
This multi-ablation structure was studied experimentally by \cite{fujioka2004} in 2004. They observed the evolution of a DAF structure inside a CHBr target and showed that this structure leads to the suppression of the Rayleigh-Taylor instability, because of the larger ablation velocity and the longer density scale length. This structure was also produced more recently at the OMEGA laser by \cite{Hager2013}.

\section{Experimental aspects}
\begin{figure}
\begin{center}
\includegraphics[scale=0.40]{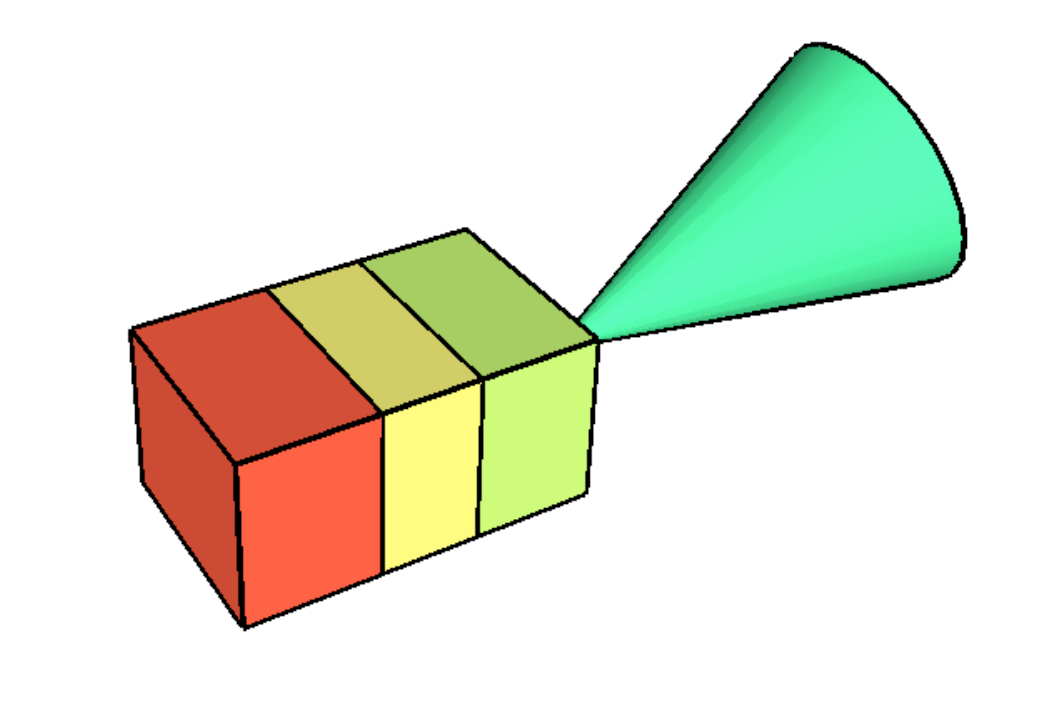}
\includegraphics[scale=0.45]{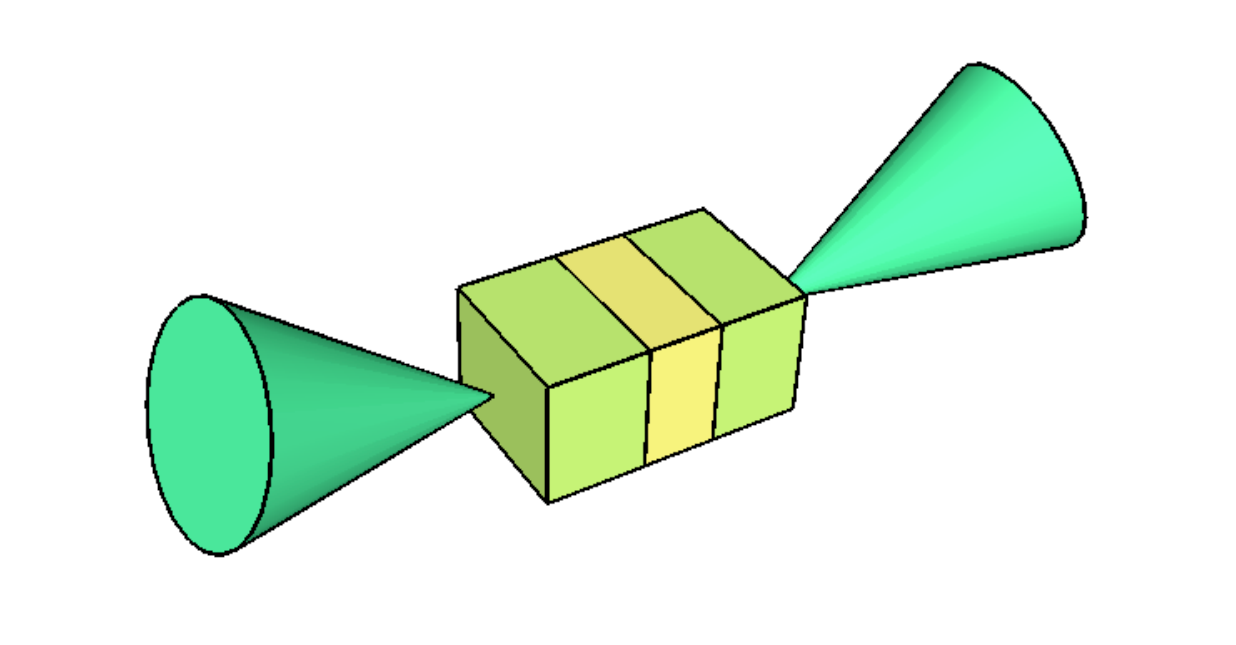}
\caption{\label{fig:epsart} Left : One-side irradiation target, composed of one layer of ablator (green), one layer of the sample of interest (yellow) and one layer of tamper (red). Right : symmetrical irradiation target, composed of two layers of ablators (green) tampering the sample of interest (yellow) (Color online).}
\label{fig6}
\end{center}
\end{figure}

We performed simulations on three-layer planar targets with two different designs: the first type is constituted of one layer of ablator, creating the DAF structure, followed by the sample of interest, tampered by a layer of CH and irradiated by only one laser beam; the second design is constituted of two layers of ablator, tampering the sample of interest and attacked symmetrically by two laser beams (Figure \ref{fig6}). 

The objective is to measure the transmission of X-rays through the sample that is related to the opacity by:
\begin{center}
$T(\nu)$=$e^{-\kappa(\nu)\rho r}$
\end{center} 
where $\rho$ is the density, $\kappa$ the spectral opacity and r the thickness of the probed plasma. \\
\noindent
 We extract from the simulation the mean density $\rho_{mean}$, temperature  $T_{e_{mean}}$ and free electron density $N_{e_{mean}}$, defined by: 
$$\rho_{mean} = \frac{\int{\rho dm}}{\int{dm}} \;\;\;
T_{e_{mean}}=\frac{\int{T_{e} dm}}{\int{dm}} \;\;\;
N_{e_{mean}} = \frac{\int{N_e dm}}{\int{dm}}$$
with their gradients on the plateau and in the iron foil, $dm$ is the mass element.

\section{Measurements of iron absorption spectra}%-------------------------------------------------------------------------------------------------------------------------------------------------------------------

\paragraph{Choice of the ablator}
We have investigated different materials to find the best one for heating iron and oxygen at the solar conditions.  We performed simulations with the same target design, modifying only the type of the ablator. We use a laser intensity I$_{laser}$=1 $\times$ 10$^{15}$ W/cm$^{2}$, irradiating three-layer targets: ablator (10{~}$\mu$m) / Fe (1{~}$\mu$m) / CH (10{~}$\mu$m) and one follows the evolution of the mean density, the mean temperature and the mean free electron density in the sample (see Table \ref{tab2}) . 
\begin{table}[h]
\begin{center}
\caption{Temperature, density and free electrons density in an iron sample during the plateau region obtained with different ablators and a laser irradiation of 1 $\times$ 10$^{15}$ W/cm$^{2}$.}
\begin{tabular}{c|c|c|c|c}
\hline
Ablator & Z & T (eV) & $\rho$ (g/cm$^{-3}$) & N$_e$ (cm$^{-3}$)\\
\hline
\hline
Titanium & 22 & 150 & 0.50 & 0.6 $\times$ 10$^{23}$\\

Copper & 29 & 130 & 0.25 & 1 $\times$ 10$^{23}$\\

Silicon & 14 & 155 & 0.90 & 1.4 $\times$ 10$^{23}$ \\

Sapphire & $\sim$14.8 & 140 & 1.16 & 1.7 $\times$ 10$^{23}$\\

Quartz & 10 & 153 & 1.14 & 1.6 $\times$ 10$^{23}$\\

Aluminium & 13 & 155 & 0.85 & 1.3 $\times$ 10$^{23}$\\
\label{tab2}
\end{tabular}
\end{center}
\end{table} 
We note that  high Z ablators (Z $>$ 20) lead to important radiative effects (high conversion between the laser energy and the X-rays). Consequently, the sample is strongly pre-heated and strongly expands. When the shock goes through the sample, the increase of the density is hence less important in comparison with a moderated Z ablator. The density reached during the plateau is then smaller (in the case of titanium (Z = 22): 0.5 g/cm$^{3}$; in the case of quartz (Z = 10): 1.14 g/cm$^{3}$).

\begin{figure}[h]
\begin{center}
\includegraphics[scale=0.19]{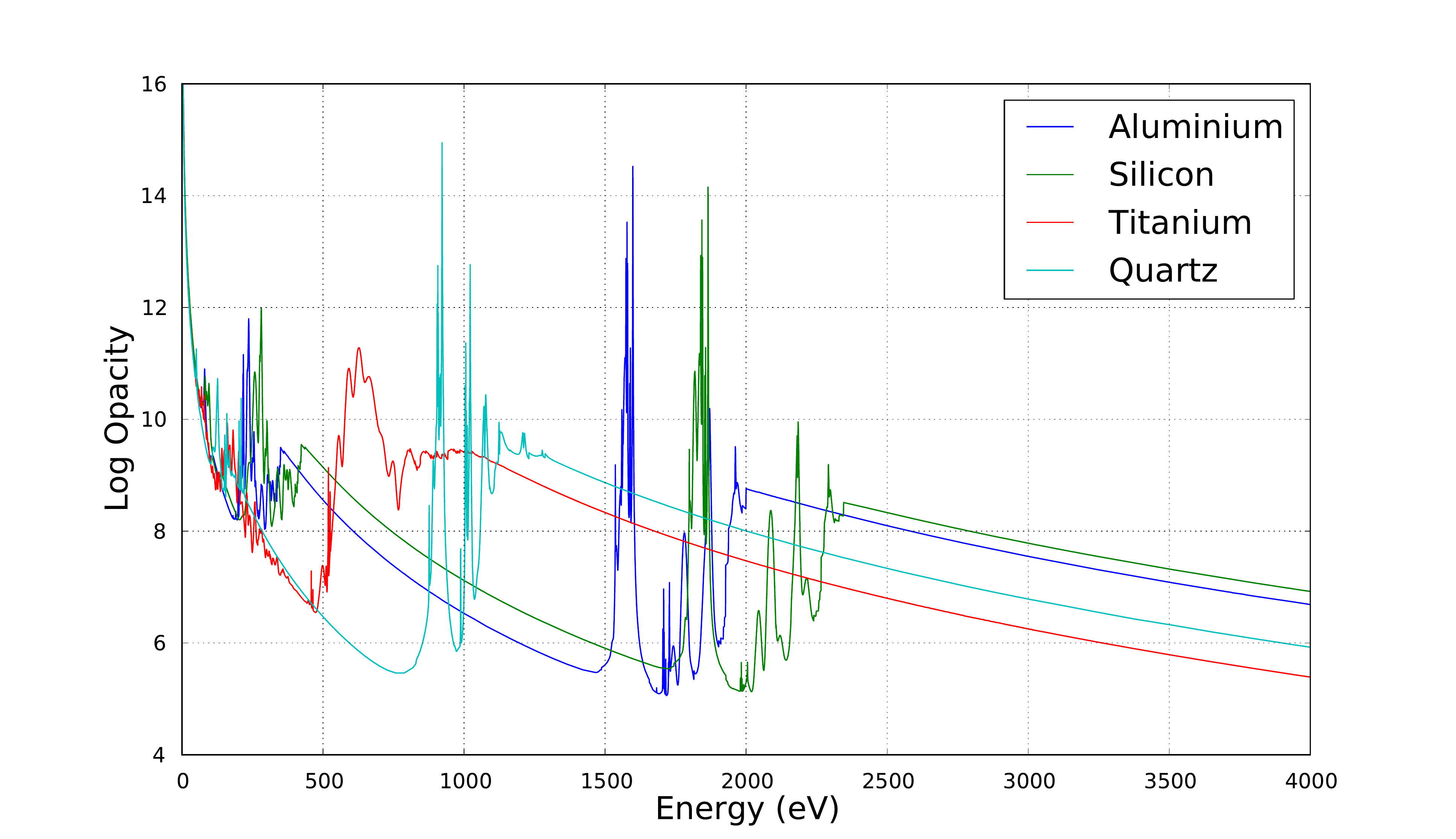}\\
\caption{\label{fig:epsart} Opacity of the considered ablators, in typical conditions of density and temperature of the simulated experiment (at the transmission measurement time) (Color online).}
\label{fig7}
\end{center}
\end{figure}

To discriminate between the different ablators, the spectra of the transmitted photons have to be considered. The objective is to avoid perturbation in the energy range of interest. 
Figure \ref{fig7} shows the comparison of the opacity of the different materials during the plateau region in the sample. Silicon is the best adapted ablator as it does not perturbed the region between 500 and 1500 eV where the important structures that we would like to study are present.
So the following simulations use silicon as an ablator.

\paragraph{Irradiation with one beam}

For a Si (8 $\mu$m) / Fe (0.1 $\mu$m) / CH (7 $\mu$m) target, irradiated by a laser intensity of 1.5 $ \times$ 10$^{15}$ W/cm$^{2}$, we obtain 0.75 $\rm  < \rho_{mean} < $ 1 g/cm$^{3}$, 160  $\rm  <  T_{mean}  < $ 180 eV and 1.1 $\rm  <  N_{e_{mean}} < $ 1.5 $\times$ 10$^{23}$ cm$^{-3}$ in the sample on a plateau of 0.7 ns.
The spatial gradients are no more than 8 \% in the sample.

%-------------------------------------------------------------------------------------------------------------------------------------------------------------------
\paragraph{Irradiation with two beams}

\begin{figure}[h]
\begin{center}
 \includegraphics[scale=0.30]{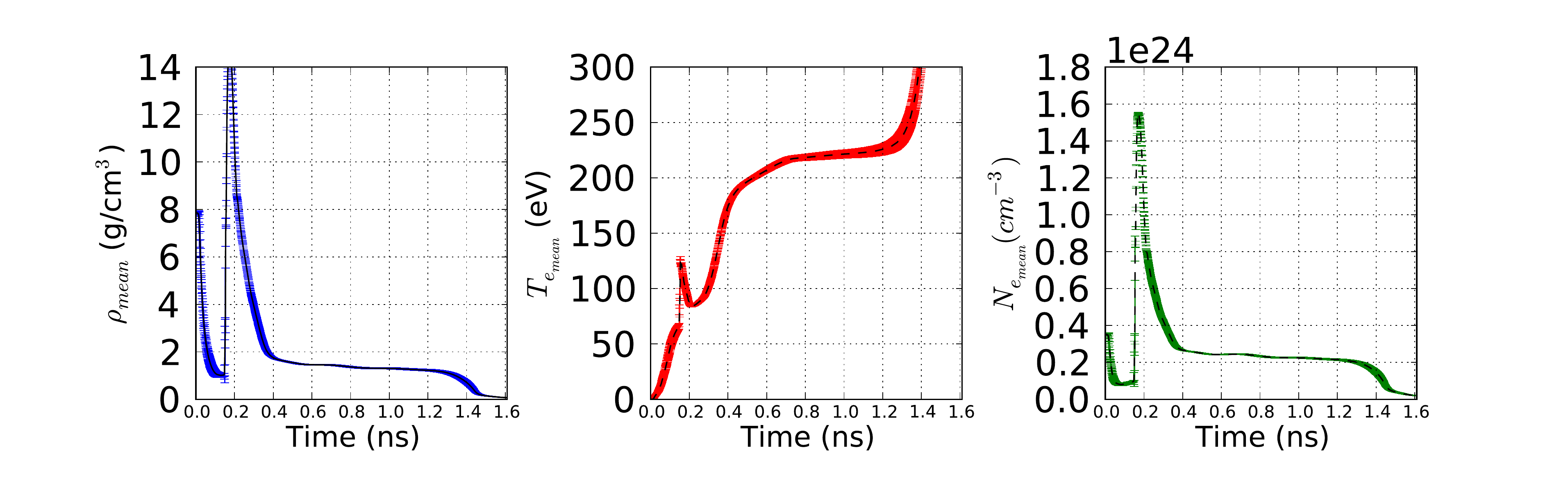} 
 \caption{Mean density, temperature and free electron density profiles obtained with a symmetrical laser irradiation of intensity 1.5 $\times$ 10$^{15}$ W/cm$^{2}$ on Si (7 $\mu$m) / Fe (0.1 $\mu$m) / Si (7 $\mu$m) (Color online).}
   \label{fig8}
   \end{center}
\end{figure}

In the case of a symmetrical irradiation by two beams of same intensity 1.5 $\times$ 10$^{15}$ W/cm$^{2}$ arriving on a target of Si (7 $\mu$m) / Fe (0.1 $\mu$m) / Si (7 $\mu$m), we got a density of 1.2 $\rm  <  \rho_{mean}  < $ 1.5 g/cm$^{3}$, 198 $\rm  <  T_{e_{mean}}  < $ 228 eV and between 2.2 $\rm  <  N_{e_{mean}}  < $ 2.50 $\times$ 10$^{23}$ cm$^{-3}$ with gradients lower than 5 \% on 0.7 ns in the iron sample (see Figure \ref{fig8}). 
 \begin{figure}[h!]
\begin{center}
 \includegraphics[scale=0.25]{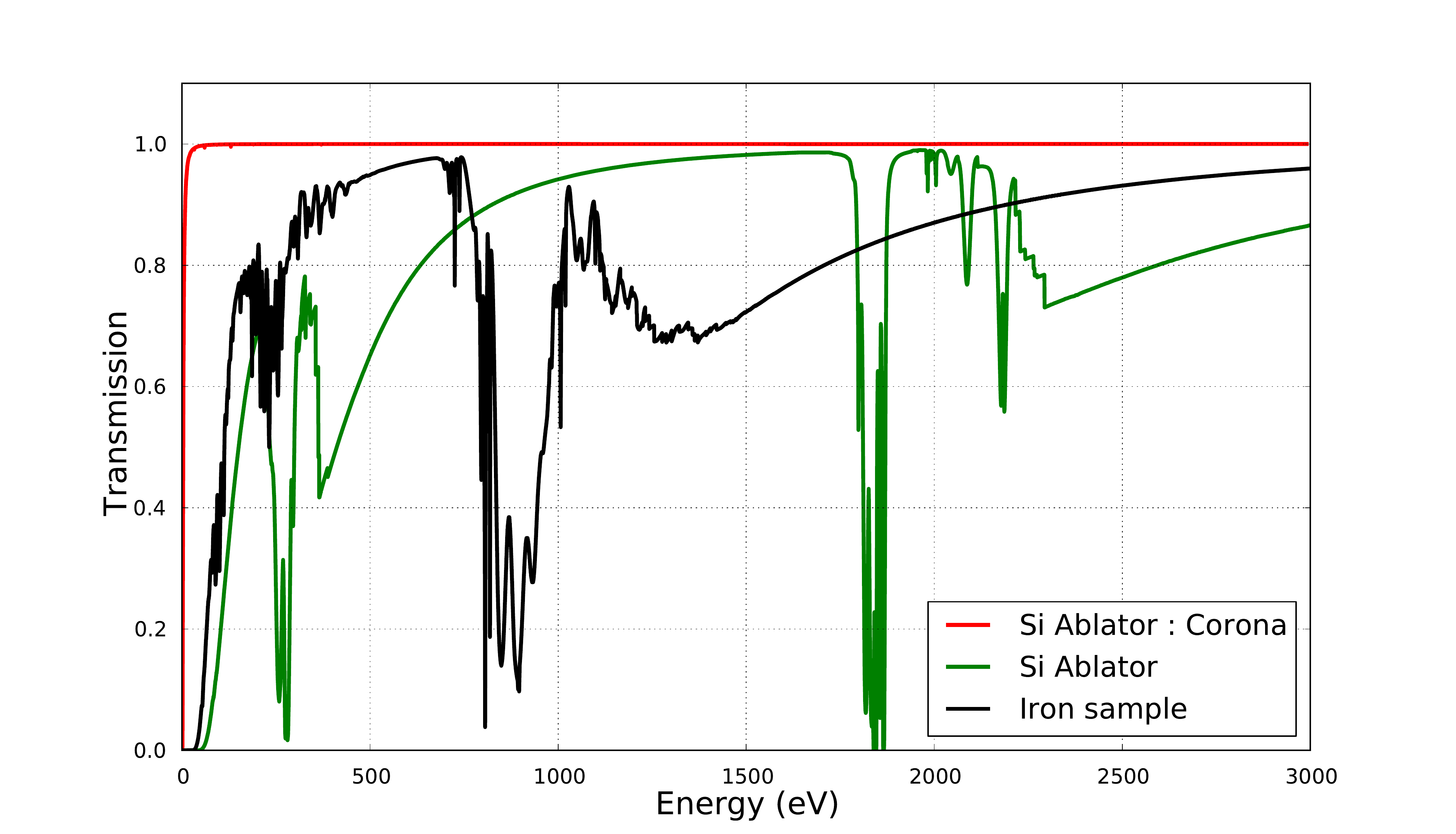} 
 \caption{Transmission of iron and silicon at the measurement time obtained with a symmetrical laser irradiation of intensity 1.5 x10$^{15}$ W/cm$^{2}$ on the Si (7 $\mu$m) / Fe (0.1 $\mu$m) / Si (7 $\mu$m) target (Color online).}
   \label{fig9}
   \end{center}
\end{figure}
\linebreak
Figure~\ref{fig9}~shows the transmission of the whole target. The choice of silicon as an ablator is validated as the structure of interest of the iron spectrum is clearly separated from silicon ones at these temperatures, which would allow a clear measurement. 
The charge state distribution at these conditions is pretty close from the solar one (around 20~\% differences on the fraction of the relevant ions (Fe XVI to Fe XIX) with the charge state distribution at 0.7 R$_\odot$), providing therefore a meaningful test of opacity calculations at these conditions.

The two previous simulations were performed with a laser intensity close to the first LMJ configuration. In the future or with NIF, one can expect to reach higher intensities. So a third simulation, a double irradiation with a laser intensity of 4{~}$\times${~}10$^{15}$ W/cm$^{2}$ with the same target as the previous case, has been also computed. 
The results of the simulation give between 2.0 and 2.3 g/cm$^{3}$ in mean density, between 265 and 290 eV in temperature and between 3.7 and 4.1 $\times$ 10$^{23}$ cm$^{-3}$ for $ \rm{N_{e_{mean}}} $. All spatial gradients were under 3 $\%$, due to the huge compression induced by the lasers on each side of the target. The charge state distribution exhibits around 4 \% of difference on the main ion (Fe XVIII) and around 15 - 20 \% for the others with the charge state distribution of iron at 0.6 R$_{\odot}$. So, with even higher intensities, one can expect to reach higher temperatures and densities, with very high stability in space and time in the sample and then might reproduce conditions corresponding to the middle of the radiative zone.

\section{Measurements of oxgen absorption spectra}
\begin{figure}[h!]
\begin{center}
 \includegraphics[scale=0.35]{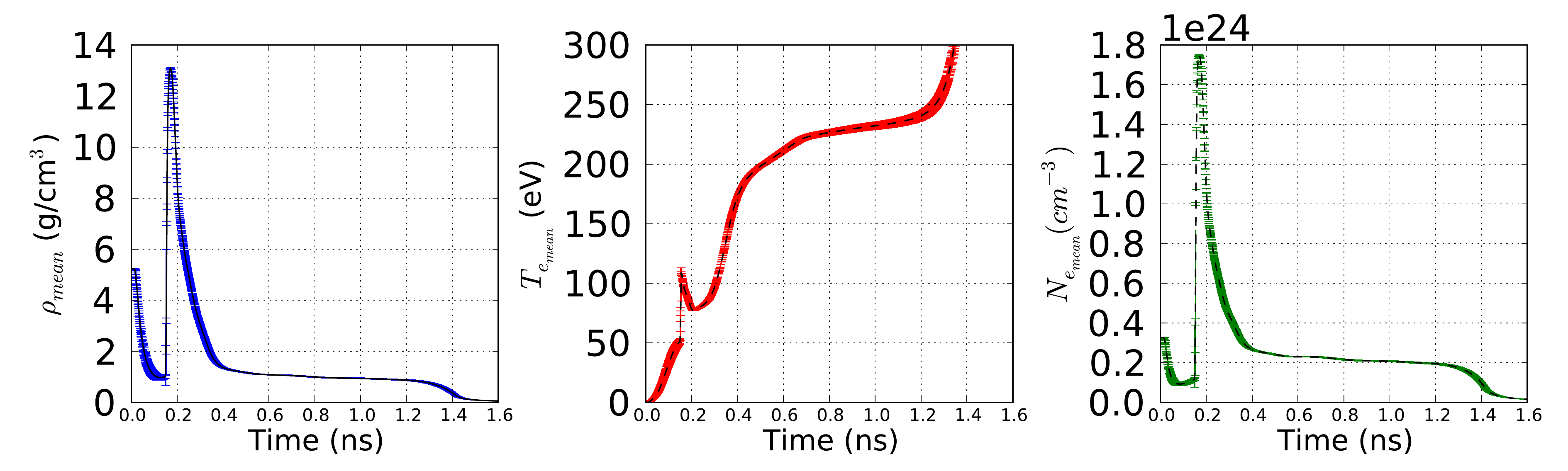} 
 \caption{Mean density, temperature and free electron density profiles obtained with a symmetrical laser irradiation of intensity 1.5 $\times$ 10$^{15}$ W/cm$^{2}$ on Si (7 $\mu$m) / Fe$_2$O$_3$ (0.1 $\mu$m) / Si (7 $\mu$m) (Color online).}
   \label{fig10}
   \end{center}
\end{figure}

\begin{figure}[h!]
\begin{center}
 \includegraphics[scale=0.25]{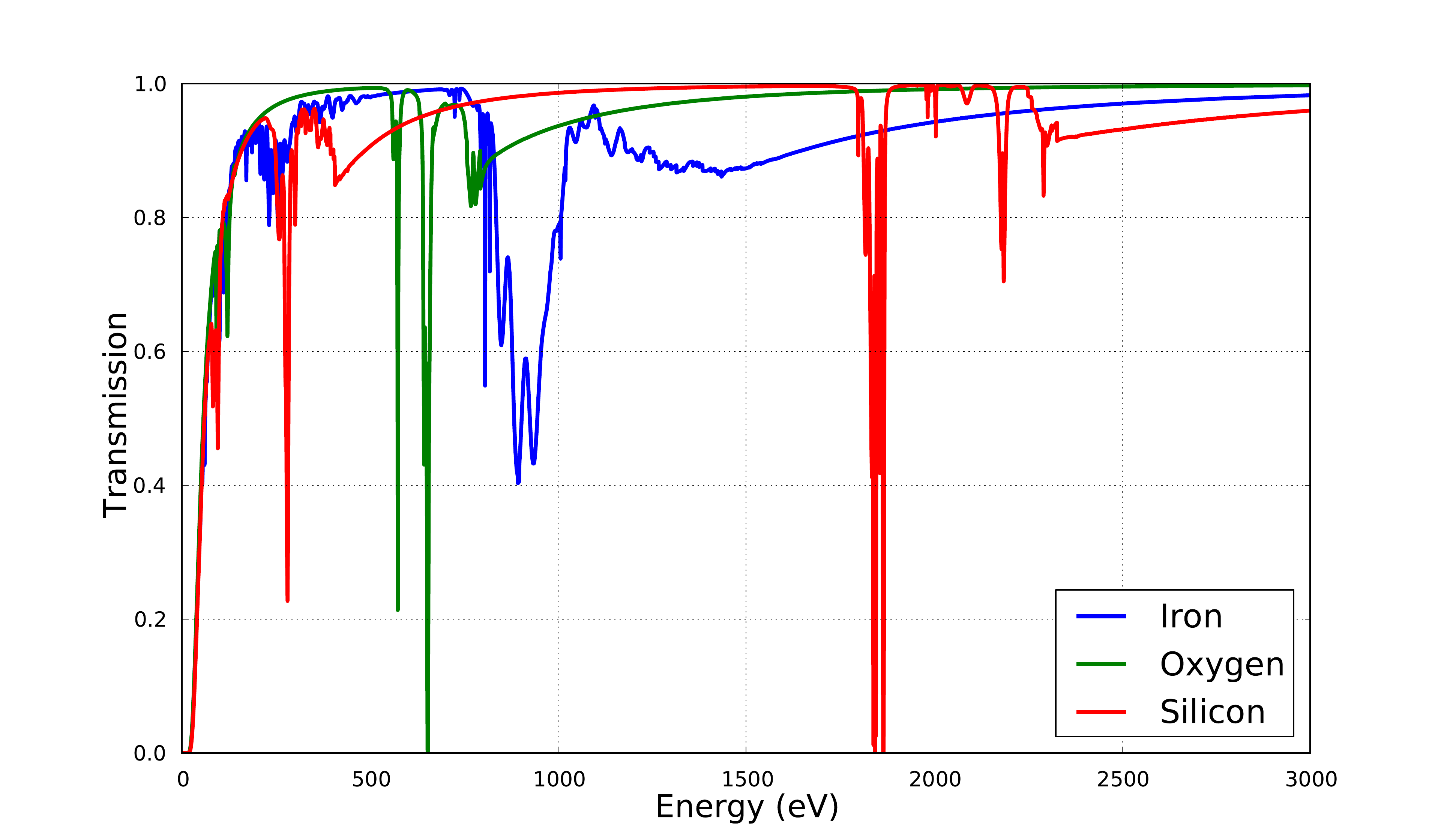} 
 \caption{Transmission of oxygen, iron and silicon in the conditions of the plateau obtained with a symmetrical laser irradiation of intensity 1.5 x10$^{15}$ W/cm$^{2}$ on the Si (7 $\mu$m) / Fe$_2$O$_3$ (0.1 $\mu$m) / Si (7 $\mu$m) target (Color online).}
   \label{fig11}
   \end{center}
\end{figure}

Oxygen is a particularly interesting element to study for solar application. Measuring the opacity of pure oxygen is really complicated at these conditions of high density and temperature : one could use a gas but then the reached density would not be high enough. That's why we propose to use oxides. We have performed simulations with the same design as the one used for iron, with hematite as a sample (Fe$_2$O$_3$): Si (7{~}$\mu$m) / Fe$_2$O$_3$ (0.1{~}$\mu$m) / Si (7{~}$\mu$m). We obtained between 0.87 and 1.08 g/cm$^{3}$ in mean density, between 211 and 242{~}eV for the mean temperature and between 1.95 and 2.3 $\times$ 10$^{23}$ cm$^{-3}$ for the free electron density (see Figure \ref{fig10}). The transmission of oxygen, iron and silicon at these conditions is represented on Figure \ref{fig11}. We are presently studying the relation between the transmission of oxygen and iron and the transmission of Fe$_2$O$_3$.

We are also performing simulations on other types of oxides to optimize the oxygen opacity measurement.
 
\section{Conclusion}
Discrepancies between the seismic observations and the prediction of the solar standard model requires an experimental validation of opacity calculations. The DAF approach creates conditions equivalent in charge state distribution and free electron density to the conditions of the radiative zone ($\mathrm{T_{e}}$ around 200 - 300 eV and $\mathrm{N_{e}}${~}$\sim${~}few{~}10$^{23}$ cm$^{-3}$), with high stability of the density and the temperature both in time and space (gradients smaller than 10 \%), convenient for LTE measurements and to check the opacity calculations for specific elements or mixture. With 1.5{~}$\times${~}10$^{15}$ W/cm$^{2}$ (corresponding to first LMJ-PETAL facility configuration), it is possible to reach interesting astrophysical conditions with high stability. With the addition of quads on LMJ-PETAL or with NIF, one will increase temperature and density to begin to look at plasma effects.\\
After the first experimental validation of this concept, the check of the line broadening will require high-resolution spectrometer for a definitive validation of the detailed calculations. This work will benefit to inertial fusion by limiting instability development in different experimental studies.

\section{Aknowledgement}
The work was supported by the ANR OPACITY. We would like to thank Y. Ralchenko for providing access to the FLYCHK code and CELIA for providing access to the CHIC code. We also thank the referee for the constructive remarks that improve the quality of the paper.

\bibliography{mybibfile}{}
\end{document}